%
%
%
%
%
%
%
%
%
%
\hoffset=0.0in
\voffset=0.0in
\hsize=6.5in
\vsize=8.9in
\normalbaselineskip=12pt
\normalbaselines
\topskip=\baselineskip
\parindent=15pt
%
%
%

\let\dl=\delta

\let\Om=\Omega

\let\lf=\left
\let\rt=\right
\let\dt=\cdot
\let\del=\nabla

\let\q=\widehat

\let\h=\hbar

\let\x=\times

\let\sy=\scriptstyle
\let\ssy=\scriptscriptstyle
\let\:=\>
\let\\=\cr
\let\emph=\e

\let\m=\hbox

\let\cl=\centerline

\def\e#1{{\it #1\/}}
\def\textbf#1{{\bf #1}}
\def\[{$$}
\def\]{\[}
\def\re#1#2{$$\matrix{#1\cr}\eqno{({\rm #2})}$$}
\def\de#1{$$\matrix{#1\cr}$$}

\def\eqdf{\buildrel{\rm def}\over =}
\def\hf{{\sy {1 \over 2}}}

\def\hfs{{\ssy {1 \over 2}}}
\def\qrs{{\ssy {1 \over 4}}}
\def\tqrs{{\ssy {3 \over 4}}}
\def\dsq{\del^2}
\def\mdsq{-\dsq}
\def\pmdsq{(\mdsq)}

\def\qOm{\q{\Om}}
\def\qOmS{\qOm_S}
\def\qOmA{\qOm_A}
\def\qOmSi{\qOm^{-1}_S}

\def\qH{\q{H}}

\def\mathrm#1{{\rm #1}}
\def\mr#1{{\rm #1}}
\def\mathcal#1{{\cal #1}}

\def\HclSA{H^{\mr{class}}_{\qOmS, \qOmA}[\phi, \pi]}
\def\LclSA{L^{\mr{class}}_{\qOmS, \qOmA}[\phi, \dot\phi]}
\def\LclS{L^{\mr{class}}_{\qOmS}[\phi, \dot\phi]}

\def\mbf{\fam\bffam\tenbf}
\def\bv#1{{\mbf #1}}

\def\vp{\bv{p}}

\def\vE{\bv{E}}
\def\vB{\bv{B}}
\def\vA{\bv{A}}

\def\vz{\bv{0}}
\def\vPhi{\bv{\Phi}}

\def\vPsi{\bv{\Psi}}
\def\vPsist{\vPsi^\ast}
\def\psist{\psi^\ast}

\def\qvp{\q{\vp}}

\def\Schr{Schr\"{o}\-ding\-er}
\font\frtbf = cmbx12 scaled \magstep1
\font\twlbf = cmbx12
\font\ninbf = cmbx9
\font\svtrm = cmr17
\font\twlrm = cmr12
\font\ninrm = cmr9
\font\ghtrm = cmr8

\def\gr#1{{\ghtrm #1}}

\def\abstract#1{{\ninbf\cl{Abstract}}\medskip
\openup -0.1\baselineskip
{\ninrm\leftskip=2pc\rightskip=2pc\noindent #1\par}
\normalbaselines}
\def\sct#1{\vskip 1.33\baselineskip\noindent{\twlbf #1}\medskip}
\def\sbsct#1{\vskip 1.15\baselineskip\noindent{\bf #1}\medskip}
\def\so{\raise 0.65ex \m{\sevenrm 1}}
\def\sk{\par\vskip 0.66\baselineskip}
{\svtrm
\cl{Klein-Gordon Transformation sans Extraneous Insertions:}
\medskip
\cl{the Isomorphic Classical Complement to a Quantum System}
}
\bigskip
{\twlrm
\cl{Steven Kenneth Kauffmann\footnote{${}^\ast$}{\gr{Retired, American
Physical Society Senior Life Member, E-mail: SKKauffmann@gmail.com}}}
}
\bigskip\smallskip
\abstract{%
The historical Klein-Gordon transformation of complex-valued first-%
order in time Schr\"{o}dinger equations iterates these in a naively
straightforward way which changes them into complex-valued second-or%
der in time equations that have a plethora of extraneous solutions---%
the transformation is an operator-calculus analogue of the squaring of
both sides of an algebraic equation.  The real and imaginary parts of 
a Schr\"{o}dinger equation, however, are well known to be precisely
the dynamical equation pair of the real-valued classical Hamiltonian
functional which is numerically equal to the expectation value of that
Schr\"{o}dinger equation's Hermitian Hamiltonian operator.  The purely
real-valued second-order in time Euler-Lagrange equation of the cor%
responding classical Lagrangian functional is also isomorphic to that
Schr\"{o}dinger equation, and for symmetric Hamiltonians has exactly
the same formal appearance as the corresponding naive complex-valued
Klein-Gordon equation, but none of the latter's extraneous solutions.
These quantum Schr\"{o}dinger-equation isomorphisms to classical Eu%
ler-Lagrange equations are the technical manifestation of a key theo%
retical aspect of the principle of complementarity, one which is ele%
gantly illustrated by the isomorphic free-photon wave-function comple%
ment to the vector potential of source-free classical electrodynamics.
}

\sct{Introduction}
\noindent
It is well known that every \Schr\ equation follows from the purely
real-valued \e{classical Hamiltonian functional} which is numerically
equal to the quantum wave-vector expectation value of its Hermitian
quantum Hamiltonian operator~[1, 2].  If one then \e{further} proceeds
to obtain the purely real-valued \e{classical Lagrangian functional
which corresponds to that particular classical Hamiltonian function%
al}, its \e{ensuing} purely real-valued second-order in time \e{clas%
sical Euler-Lagrange equation} is \e{naturally isomorphic} to the
original complex-valued first-order in time \e{quantum \Schr\ equa%
tion}.  In that way the \e{goal} of Klein and Gordon~[3] \e{is univer%
sally achieved without unintentionally injecting extraneous solu%
tions}.  In fact, when the Hermitian quantum Hamiltonian operator hap%
pens to be \e{symmetric} (and therefore real-valued), the ensuing
purely \e{real-valued} classical Euler-Lagrange equation \e{has the
very same formal appearance} as the problematic \e{complex-valued}
equation that Klein and Gordon obtained by naive operator squaring on
both sides of the \Schr\ equation~[3], a process whose algebraic ana%
logue is notorious for the unintended injection of extraneous roots.

The detailed description below of the general class of \e{quantum}
\Schr\ to \e{classical} Euler-Lagrange \e{equation isomorphisms} is
the technical rendition of a key theoretical aspect of \e{the princi%
ple of complementarity}~[4], one which is elegantly illustrated by the
isomorphic free-photon wave-function complement to the vector poten%
tial of source-free classical electrodynamics~[5, 6].

\sct{Isomorphic Euler-Lagrange complements to \Schr\ equations}
\noindent
As is well known, the \Schr\ equation,
\re{
i\h\dot\psi = \qH\psi
}{1a}
can be split into real and imaginary parts~[2] by explicitly thus
splitting \e{both} its complex-valued quantum \e{wave vector} $\psi$,
\re{
\psi = (2\h)^{-\hfs}(\phi + i\pi),\enskip\hbox{where}\enskip
\phi \eqdf  (\h/2)^\hfs(\psi + \psist)\enskip\hbox{and}\enskip
\pi \eqdf -i(\h/2)^\hfs(\psi - \psist).
}{1b}
\e{and} its Hermitian quantum \e{Hamiltonian operator} $\qH$,
\re{
\qH = \h(\qOmS + i\qOmA),\enskip\hbox{where}\enskip
\qOmS \eqdf   (\qH + \qH^\ast)/(2\h) =   (\qH + \qH^T)/(2\h)\cr
                       \hbox{and}\enskip
\qOmA \eqdf -i(\qH - \qH^\ast)/(2\h) = -i(\qH - \qH^T)/(2\h),
}{1c}
from which we see that $\qOmS$ is real-valued and \e{symmetric}, while
$\qOmA$ is real-valued and \e{antisymmetric}.

When we substitute Eqs.~(1b) and (1c) into the \Schr\ Eq.~(1a), we ob%
tain,
\re{
i\dot\phi - \dot\pi = \qOmS\phi - \qOmA\pi + i(\qOmS\pi + \qOmA\phi),
}{2a}
which yields \e{two purely real-valued} first-order in time equations
\e{that together are equivalent to the \Schr} Eq.~(1a),
\re{
\dot\phi =   \qOmS\pi + \qOmA\phi,\quad
\dot\pi  = -\qOmS\phi + \qOmA\pi.
}{2b}

It turns out that Eq.~(2b) can \e{also} be obtained as the pair of dy%
namical equations of motion that follow from the real-valued \e{clas%
sical Hamiltonian functional} which is numerically equal to the wave-%
vector expectation value of the Hermitian quantum Hamiltonian opera%
tor~[2],
\re{
\HclSA \eqdf (\psist, \qH\psi) = \hf((\phi - i\pi), (\qOmS + i\qOmA)
(\phi + i\pi))\cr
       = \hf[(\phi, \qOmS\phi) + (\pi, \qOmS\pi) + 2(\pi, \qOmA\phi)]\cr
       = \hf[(\phi, \qOmS\phi) + (\pi, \qOmS\pi) - 2(\phi, \qOmA\pi)],
}{3a}
where we have made use of three identities which follow from the facts
that $\qOmS$ is \e{symmetric} and $\qOmA$ is \e{antisymmetric}.  These
identities are,
\de{
(\phi, \qOmS\pi) = (\pi, \qOmS\phi),\quad (\phi, \qOmA\phi) =
0 = (\pi, \qOmA\pi),\quad (\pi, \qOmA\phi) = -(\phi, \qOmA\pi).
}

We now proceed to obtain the pair of Hamilton's classical dynamical
equations of motion which follow from the classical Hamiltonian func%
tional $\HclSA$ of Eq.~(3a),
\re{
\dot\phi =  \dl\HclSA/\dl\pi =  \qOmS\pi + \qOmA\phi,\cr\hbox{\ }\cr
\dot\pi = -\dl\HclSA/\dl\phi = -\qOmS\phi + \qOmA\pi.
}{3b}
These equations have come out to be \e{exactly the same} as those of
Eq.~(2b), which are, in turn, \e{equivalent} to the \Schr\ Eq.~(1a).

Now to \e{convert} the Eq.~(2b) pair of first-order in time equations
for $\phi$ and $\pi$ into an Euler-Lagrange style second-order in time
equation for $\phi$ \e{alone} requires that we \e{solve} for $\pi$ in
terms of $\phi$ and $\dot\phi$.  If the real-valued symmetric operator
$\qOmS$ has the \e{inverse} $\qOmSi$, we note from the first equality
of Eq.~(2b) that this is readily done, with the result,
\re{
\pi = \qOmSi(\dot\phi - \qOmA\phi),
}{4a}
which we can then substitute into the second equality of Eq.~(2b),
producing the second-order in time equation for $\phi$ alone,
\re{
\qOmSi(\ddot\phi - \qOmA\dot\phi) = -\qOmS\phi +
                                    \qOmA\qOmSi(\dot\phi - \qOmA\phi),
}{4b}
which is more neatly rewritten as,
\re{
\qOmSi\ddot\phi - (\qOmSi\qOmA + \qOmA\qOmSi)\dot\phi +
(\qOmS + \qOmA\qOmSi\qOmA)\phi = 0,
}{4c}
an Euler-Lagrange style purely real-valued second-order in time equa%
tion for $\phi$ \e{alone}.  We take note that the sequential steps
carried out in Eqs.~(4a), (4b), and (4c) above \e{are all reversible}.
In particular, it is apparent that Eq.~(4c) is \e{equivalent} to Eq.~%
(4b), and that Eq.~(4a) is \e{equivalent} to \e{the first equality of}
Eq.~(2b).  Furthermore, putting Eq.~(4a) into Eq.~(4b), or \e{equally
as well} into Eq.~(4c), produces \e{the second equality of} Eq.~(2b).
Therefore, \e{not only} does Eq.~(2b) \e{imply} Eqs.~(4a) and (4c),
Eqs.~(4a) and (4c) \e{also imply} Eq.~(2b), which is \e{equivalent} to
the \Schr\ Eq.~(1a).

Thus Eqs.~(4a) and (4c) above are equivalent to the \Schr\ Eq.~(1a).
This observation, however, is an infelicitous one as it stands because
Eq.~(4a) refers to $\pi$, whereas the \Schr\ Eq.~(1a) is exclusively
concerned with $\psi$.  However, \e{combining} Eq.~(4a) \e{with the
two} Eq.~(1b) \e{definitions of} $\pi$ \e{and} $\phi$ \e{in terms of}
$\psi$ \e{and} $\psist$ in the form,
\re{
-i(\h/2)^\hfs(\psi - \psist) = \qOmSi(\dot\phi - \qOmA\phi),\quad
(\h/2)^\hfs(\psi + \psist) = \phi,
}{4d}
does enable the \e{integration} of Eq.~(4a) into a formula for $\psi$
\e{alone} in terms of $\phi$ and $\dot\phi$.  This is achieved  by
linearly combining the two equalities of Eq.~(4d) so as to eliminate
$\psist$ and give $\psi$ the coefficient of unity,
\re{
\psi = (2\h)^{-\hfs}[(1 - i\qOmSi\qOmA)\phi + i\qOmSi\dot\phi].
}{4e}
It is therefore clear as a matter of mathematics that Eq.~(4e) will,
together with the Euler-Lagrange style Eq.~(4c), permit derivation of
the \Schr\ Eq.~(1a).  That fact \e{can also be directly verified} by
making use of the ``left factorization identities'',
\de{
(\qOmS + \qOmA\qOmSi\qOmA) = (\qOmS + i\qOmA)(1 - i\qOmSi\qOmA)
\quad\hbox{and}\quad(1 + i\qOmA\qOmSi) = (\qOmS + i\qOmA)\qOmSi,
}
\e{after} differentiating both sides of Eq.~(4e) with respect to time
and then inserting Eq.~(4c) into the right-hand side of the result of
that differentiation.

Of course the steps which we have displayed up to this point \e{also
make it apparent} that Eq.~(4e) and the Euler-Lagrange style Eq.~(4c)
\e{both follow from} the \Schr\ Eq.~(1a) and the \e{definitions and
properties} of $\phi$, $\qOmS$ and $\qOmA$ that are set out in Eqs.~%
(1b) and (1c).  Therefore, if the real-valued symmetric operator
$\qOmS$ indeed has the inverse $\qOmSi$, then the relation of the Eu%
ler-Lagrange style Eq.~(4c) to the \Schr\ Eq.~(1a) \e{is an isomorphic
one via} Eq.~(4e).

Eq.~(4e) is, of course, \e{readily inverted} after the extraction of
its real and imaginary parts, with the result,
\re{
\phi =       (\h/2)^\hfs(\psi + \psist),\cr\hbox{\ }\cr
\dot\phi = -i(\h/2)^\hfs[(\qOmS + i\qOmA)\psi - (\qOmS - i\qOmA)
\psist].
}{4f}
Of course the first equality of Eq.~(4f) is nothing more than the
\e{definition} of $\phi$ which is given in Eq.~(1b), while the second
equality of Eq.~(4f) simply follows from straightforward application
of the \Schr\ Eq.~(1a) to the first equality.

In fact if we then proceed to \e{likewise} apply the \Schr\ Eq.~(1a)
to the \e{second} equality of Eq.~(4f), and then use Eq.~(4e) to
\e{eliminate} $\psi$ and $\psist$ from the right hand side of what re%
sults, we find that (after a great many cancellations of multi-factor
terms) the Euler-Lagrange style Eq.~(4c) emerges.

In a nutshell, there are clearly many different ways (of varying alge%
braic complexity) to establish the \e{isomorphism} of the Euler-La%
grange style Eq.~(4c) to the \Schr\ Eq.~(1a) via the relationship ex%
pressed by Eq.~(4e) (or by its inverse Eq.~(4f)).

In the not infrequently occurring special circumstance that the Hermi%
tian Hamiltonian operator $\qH$ is \e{symmetric} (and therefore real-%
valued), the real-valued antisymmetric operator $\qOmA$ \e{vanishes
identically}, which drastically simplifies the purely real-valued Eu%
ler-Lagrange style Eq.~(4c).  That equation thereupon becomes,
\re{
\ddot\phi + (\qOmS)^2\phi = 0\quad\hbox{or}\quad\ddot\phi +
(\qH/\h)^2\phi = 0,
}{4g}
Eq.~(4g) \e{has the very same formal appearance} as the equation for
the \e{complex-valued} quantum wave vector $\psi$ which was proposed
by Klein and Gordon~[3].  In Eq.~(4g), however, $\phi$ is (within a
real-valued constant) \e{only the real part of} $\psi$, as is com%
pletely apparent from Eq.~(1b).  This \e{purely real-valued property
of} $\phi$ is obviously a \e{critical aspect} of the \e{isomorphic re%
lationship} of the \Schr\ Eq.~(1a) to Eq.~(4c), which in this particu%
lar circumstance has reduced to Eq.~(4g).  In short, the complex-val%
ued version of Eq.~(4g) proposed by Klein and Gordon, wherein the
\e{real-valued} $\phi$ is replaced by the \e{complex-valued} $\psi$,
obviously \e{destroys} the isomorphic relationship of Eq.~(4g) to the
\Schr\ Eq.~(1a) by \e{doubling} the number of its effective degrees of
freedom relative to those of the \Schr\ equation, \e{which obviously
injects a plethora of extraneous solutions}.

Taking up a loose end now which hasn't yet been discussed, in case
that $\qOmS$ \e{doesn't actually have an inverse}, then \e{for a phys%
ically sensible system} we would \e{nevertheless} expect the eigenval%
ue spectrum of \e{both} $\qH$ \e{and} $\qOmS$ to be \e{bounded below}.
So for a physically sensible system there ought to exist a nonnegative
energy constant $E_0$ such that \e{both} $\qH + E_0$ \e{and} $\qOmS +
(E_0/\h)$ \e{do have inverses}.  Since the \e{constants} $E_0$ and
$(E_0/\h)$ \e{commute with all operators}, the thus \e{modified} quan%
tum model \e{will differ only almost trivially} from the original one:
one merely needs to \e{subtract the constant value} $E_0$ from all en%
ergy eigenvalues once the calculation is completed, while as well
\e{multiplying} any \e{time-evolved} complex-valued quantum wave vec%
tor $\psi(t; t_0)$ \e{by the simple phase factor} $\exp[i(E_0/\h)(t -
t_0)]$.

Finally, we wish to formally complete this discussion by explicitly
passing from from the classical Hamiltonian functional $\HclSA$ of
Eq.~(3a) \e{to its corresponding classical Lagrangian functional}
$\LclSA$, and then obtain from the latter the formally resulting pure%
ly real-valued second-order in time Euler-Lagrange classical dynamical
equation of motion for $\phi$ alone.  Since that Euler-Lagrange clas%
sical dynamical equation of motion must necessarily be \e{consistent}
with the \e{corresponding classical Hamiltion} pair of first-order in
time dynamical equations of motion that are given by Eq.~(3b), an
equation pair which is \e{identical} to Eq.~(2b), which is in turn a
\e{transcription} of the real and imaginary parts of the \Schr\ Eq.~%
(1a), \e{there is simply no way} that that Euler-Lagrange equation of
motion \e{could differ} from the Euler-Lagrange style Eq.~(4c), which
is indeed \e{isomorphic} to both the \Schr\ Eq.~(1a) as well as to the
Eq.~(2b) transcription of its real and imaginary parts.  In fact, Eq.%
~(4c) \e{was specifically developed} to be \e{equivalent} to the sec%
ond equality of Eq.~(2b) by way of the first equality of Eq.~(2b).

Although there is thus \e{no question} that Eq.~(4c) is the \e{cor%
rect} Euler-Lagrange classical equation of motion that is \e{isomor%
phic} to the \Schr\ Eq.~(1a), it is still of \e{technical} interest
to \e{explicitly} work out the \e{classical Lagrangian functional}
$\LclSA$ which is in fact \e{complementary} to the Eq.~(1a) \e{quantum
\Schr\ system}.

Passage from the Eq.~(3a) classical Hamiltonian functional of $\phi$
and $\pi$ to its corresponding classical Lagrangian functional of
$\phi$ and $\dot\phi$ of course requires that we \e{solve} for $\pi$
in terms of $\phi$ and $\dot\phi$.  But \e{exactly that solution} is
given by Eq.~(4a) above.  What remains to obtain the classical Lagran%
gian functional $\LclSA$ is largely algebra, albeit rather burdensome
in its amount,
\re{
\LclSA = (\dot\phi, \pi) -
\lf.\HclSA\rt|_{\pi = \qOmSi(\dot\phi - \qOmA\phi)}\cr\hbox{\ }\cr
       = (\dot\phi, \qOmSi(\dot\phi - \qOmA\phi)) - \hf(\phi, \qOmS\phi)
-\hf(\qOmSi(\dot\phi - \qOmA\phi), (\dot\phi - \qOmA\phi))
+(\phi, \qOmA\qOmSi(\dot\phi - \qOmA\phi))\cr\hbox{\ }\cr
       = \hf[(\dot\phi, \qOmSi\dot\phi) - 2(\dot\phi, \qOmSi\qOmA\phi) -
            (\phi, (\qOmS + \qOmA\qOmSi\qOmA)\phi)]\cr
       = \hf[(\dot\phi, \qOmSi\dot\phi) + 2(\phi, \qOmA\qOmSi\dot\phi) -
            (\phi, (\qOmS + \qOmA\qOmSi\qOmA)\phi)],
}{5a}
where \e{multiply repeated} use has been made of the \e{symmetric} na%
ture of $\qOmSi$ and the \e{antisymmetric} nature of $\qOmA$.

We can now functionally differentiate $\LclSA$ with respect to both
$\dot\phi$ and $\phi$,
\re{
\dl\LclSA/\dl\dot\phi = \qOmSi\dot\phi - \qOmSi\qOmA\phi,\cr\hbox{\ }\cr
\dl\LclSA/\dl\phi = \qOmA\qOmSi\dot\phi - (\qOmS + \qOmA\qOmSi\qOmA)\phi.
}{5b}
Therefore, the Euler-Lagrange equation,
\re{
d(\dl\LclSA/\dl\dot\phi)/dt = \dl\LclSA/\dl\phi,
}{5c}
works out to,
\re{
\qOmSi\ddot\phi - (\qOmSi\qOmA + \qOmA\qOmSi)\dot\phi +
(\qOmS + \qOmA\qOmSi\qOmA)\phi = 0,
}{5d}
namely exactly the same as Eq.~(4c), which is definitely expected.

In the circumstance that $\qH$ happens to be \e{symmetric} (and there%
fore real-valued), $\qOmA$ vanishes identically and $\LclSA$ greatly
simplifies to become,
\re{
\LclS = \hf[(\dot\phi, \qOmSi\dot\phi) - (\phi, \qOmS\phi)],
}{5e}
whose correspondingly simplified real-valued Euler-Lagrange equation
is of course given by Eq.~(4g).

In this circumstance that $\qH$ is symmetric, $\qH = \h\qOmS$ from
Eq.~(1c), and the \Schr\ Eq.~(1a) is simply written as,
\re{
\dot\psi = -i\qOmS\psi\quad\hbox{or}\quad\dot\psist = i\qOmS\psist,
}{6a}
from which we can see Eq.~(4f) simplifies to,
\re{
\phi = (\h/2)^\hfs(\psi + \psist),\quad\enskip
\dot\phi = -i(\h/2)^\hfs\qOmS(\psi - \psist),
}{6b}
whose inverse is also much simpler than Eq.~(4e),
\re{
\psi = (2\h)^{-\hfs}(\phi + i\qOmSi\dot\phi),
}{6c}
and whose Euler-Lagrange Eq.~(4g) follows from application of the
\Schr\ Eq.~(6a) to the second equality of Eq.~(6b),
\re{
\ddot\phi + (\qOmS)^2\phi = 0\quad\hbox{or}\quad\ddot\phi +
           (\qH/\h)^2\phi = 0,
}{6d}

Since source-free classical electrodynamics in radiation gauge obeys
the classical wave equation, which has the \e{form} of the simplified
Euler-Lagrange Eq.~(6d), we can immediately deduce from it the \e{ap%
plicable} real-valued symmetric operator $\qOmS$, and therefore \e{the
nature of its complementary isomorphic \Schr\ equation}, which, as
pointed out in the first sentence of the foregoing paragraph, has the
Hamiltonian operator $\qH = \h\qOmS$.

Is is instructive and lamentable to note that notwithstanding the work
of Klein and Gordon with quantum-mechanics related equations of the
form of Eq.~(6d), the isomorphic quantum complement to source-free
classical electrodynamics which we are about to discuss \e{was a com%
pletely closed book} for those theorists because the equations of
classical electrodynamics are, of course, \e{strictly real-valued},
whereas Klein and Gordon insisted on considering Eq.~(6d) forms \e{on%
ly in conjunction with strictly complex-valued solutions}~[3].  In this we
have a worthwhile reminder that physics confronts beleaguered theo%
rists with a multitude of subtle ways to wander astray.  Still, it can
reasonably be contended that Klein and Gordon \e{ought} to have re%
garded the surfeit of \e{extraneous solutions} which they soon encoun%
tered as a stern warning to backtrack and rethink.

\sbsct{The isomorphic quantum complement to source-free classical
electrodynamics}
\noindent
Source-free classical electrodynamics in radiation gauge has vanishing
scalar potential, while its vector potential is transverse and satis%
fies the classical wave equation~[7], i.e.,
\re{
A^0 = 0,\quad \del\dt\vA = 0,\quad \ddot\vA - c^2\dsq\vA = \vz.
}{7a}
The classical wave equation certainly has the special simplified Eul%
er-Lagrange form set out in Eq.~(6d) with $\qOmS = c\pmdsq^\hfs$ or
$\qH = \h c\pmdsq^\hfs$.  Now in quantum mechanical configuration rep%
resentation the three-momentum operator $\qvp$ for a particle is equal
to $-i\h\del$, so that the quantum Hamiltonian operator $\qH = \h c
\pmdsq^\hfs$ which we have just deduced from the classical wave equa%
tion corresponds in particle three-momentum quantum operator terms to
$\qH = c|\qvp|$, which is the relativistic quantum Hamiltonian opera%
tor of a \e{zero rest-mass free particle} that has the three-momentum
quantum operator $\qvp$.  In other words, it appears that the quantum
complement of classical source-free electrodynamics is the \Schr\
equation \e{of the free photon}, which, of course has zero rest mass.

\e{Further} information about the \e{complex-valued} photon wave func%
tion ought to be available from Eq.~(6c) above.  However, \e{because}
the source-free \e{classical} electrodynamics potential $\vA$ is a
\e{vector} field, it would naturally be expected that Eq.~(6c) would
be slightly \e{modified} to assume the complex-valued \e{vector} wave
function form,
\re{
\vPsi = (2\h)^{-\hfs}(\vPhi + (i/c)\pmdsq^{-\hfs}\dot\vPhi).
}{7b}
We obviously \e{hope} to identify the real-valued vector field $\vPhi$
as just the radiation gauge classical vector potential $\vA$ \e{it%
self}.  However, if we require \e{the dimension of} $\vPsi$ \e{to be
that of a quantum mechanical wave function}, the identification of
$\vPhi$ as $\vA$ itself \e{isn't dimensionally feasible}.  That minor
dimensional impasse is \e{readily resolved}, however, by inserting di%
mensionally needed bits and pieces of \e{already known physical fac%
tors that actually pertain to this classical electrodynamic system},
namely by letting,
\re{
\vPhi\eqdf c^{-\hfs}\pmdsq^\qrs\vA,
}{7c}
so that Eq.~(7b) becomes,
\re{
\vPsi = (2\h c)^{-\hfs}(\pmdsq^\qrs\vA + (i/c)\pmdsq^{-\qrs}\dot\vA),
}{7d}
and $\vPsi$ is readily verified to have the correct dimension for a
quantum mechanical wave function.

From the classical wave equation, namely the third equality of Eq.~%
(7a), we have that $\ddot\vA = -c^2\pmdsq\vA$.  Using this, we obtain
from Eq.~(7d) (after performing some algebra) that,
\re{
\dot\vPsi = -ic\pmdsq^\hfs\vPsi,
}{7e}
which is consistent with the \Schr\ equation for the free photon be%
cause the free-photon Hamiltonian operator is $\qH = \h c\pmdsq^\hfs =
c|\qvp|$.

In \e{addition} to this deduction of the \Schr\ equation for the free
photon, we can use the \e{second equality} of Eq.~(7a), namely that
$\del\dt\vA = 0$, to deduce from Eq.~(7d) that,
\re{
\del\dt\vPsi = 0.
}{7f}
Therefore the free-photon complex-valued \e{vector} wave function is
\e{transverse}, which implies that the free photon has \e{transverse}
polarization (i.e., \e{transverse} ``spin'').

Finally, it will be feasible to \e{invert} Eq.~(7d), obtaining the
\e{classical} $\vA$ and $\dot\vA$ in terms of the \e{quantum} $\vPsi$
and $\vPsist$, from which the classical wave equation can be obtained
as a \e{consequence} of the \Schr\ equation, e.g., Eq.~(7e).  This is
the \e{complement} of what was established when the effective \Schr\
Eq.~(7e) was obtained through use on Eq.~(7d) of the \e{classical}
wave equation.  That the quantum and classical equations \e{in fact
imply each other} is, of course, the \e{heart} of the \e{complementary
isomorphism}.

Adding Eq.~(7d) to its complex conjugate, and subtracting its complex
conjugate from it permits us to obtain its inversion, which is,
\re{
\vA = ((\h c)/2)^\hfs  \pmdsq^{-\qrs}(\vPsi + \vPsist),\cr\hbox{\ }\cr
\dot\vA = -ic ((\h c)/2)^\hfs  \pmdsq^\qrs(\vPsi - \vPsist).
}{7g}
Note that, as is usual is these cases, the second equality of Eq.~(7g)
follows from the first by application to it of the \Schr\ equation,
e.g., $\dot\vA$ can be obtained through application of Eq.~(7e) to
$\vPsi$ and $\vPsist$ on the right-hand side of the first equality of
Eq.~(7g).

It is now \e{furthermore} readily seen from Eqs.~(7g) and (7e) that
application of the \Schr\ equation to the \e{second} equality of Eq.%
~(7g) proves to be \e{equivalent} to the \e{classical wave equation
for} $\vA$, i.e., to the third equality of Eq.~(7a).  In \e{addition},
application of Eq.~(7f) to the first equality of Eq.~(7g) shows that
$\del\dt\vA = 0$, i.e., it demonstrates the \e{second} equality of
Eq.~(7a).  Therefore the \e{quantum} Eqs.~(7e) and (7f) for $\vPsi$
imply the two \e{classical} Eq.~(7a) equalities for $\vA$.  Those im%
plications are, of course, the needed \e{second part} of establishing
the \e{complementary isomorphism} of the classical equations for $\vA$
to the quantum equations for $\vPsi$, as is pointed out in the para%
graph which precedes the one immediately above.

A last fact which is of interest is that it is feasible \e{swap out}
of Eqs. (7d) and (7g) the source-free radiation gauge electromagnetic
\e{potential} $\vA$ described by the three equalities of Eq.~(7a)
\e{in favor} of the source-free \e{electric and magnetic fields}.
Those source-free electric and magnetic fields are governed by the
four well-known source-free classical Maxwell field equations,
\re{
\dot\vB = -c\del\x\vE,\quad \dot\vE = c\del\x\vB,\quad
\del\dt\vB = 0,\quad  \del\dt\vE = 0.
}{8a}
It turns out that these four source-free classical Maxwell equations
\e{not only imply} the free-photon effective \Schr\ Eq.~(7e) and the
free-photon transverse-spin condition of Eq.~(7f), they are as well
\e{in turn implied by} those free-photon quantum requirements.  Thus
we \e{also} have a situation of perfect \e{complementary isomorphism}
of the physical behavior of the quantum free photon to the \e{complete
set} of source-free classical Maxwell equations.  Faraday and Maxwell
were therefore the first physicists to practice quantum mechanics---%
indeed on the ultra-relativistic free photon and its peculiar trans%
verse spin at that.

To pass from the source-free radiation gauge potential of Eq.~(7a) to
the source-free electric and magnetic fields, we note that the first
Eq.~(7a) equality, namely $A^0 = 0$, implies that $\vE = -\dot\vA/c$,
or $\dot\vA = -c\vE$ .  Also the fact that $\vB = \del\x\vA$, together
with the second Eq.~(7a) equality, namely $\del\dt\vA = 0$, implies
that $\del\x\vB = \mdsq\vA$, or $\vA = \pmdsq^{-1}(\del\x\vB)$.  With
these relations in hand, we reexpress Eq.~(7d) in terms of the source%
-free $\vE$ and $\vB$ fields,
\re{
\vPsi = (2\h c)^{-\hfs}(\pmdsq^{-\tqrs}(\del\x\vB) - i\pmdsq^{-\qrs}
\vE),
}{8b}
From the particular Eq.~(8a) source-free classical Maxwell equation
$\del\dt\vE = 0$, Eq.~(8b) immeditately yields,
\re{
\del\dt\vPsi = 0,
}{8c}
which is exactly the same free-photon transverse spin requirement as
we previously obtained in the form of Eq.~(7f).  Furthermore, since
it can be straightforwardly shown that the four Eq.~(8a) source-free
Maxwell equations imply both of the relations $\dot\vE = c(\del\x\vB)$
and $\del\x\dot\vB = -c\pmdsq\vE$, we \e{also} obtain from Eq.~(8b)
(after performing some algebra) that,
\re{
\dot\vPsi = -ic\pmdsq^\hfs\vPsi,
}{8d}
which is exactly the same effective \Schr\ equation for a free photon
as we previously obtained in the form of Eq.~(7e).

If we extract the real and imaginary parts of Eq.~(8b), it turns out
that we can readily solve those for the $\vB$ and $\vE$ fields, ex%
pressed, respectively, in terms of the real and imaginary parts of
the complex-valued free-photon wave function $\vPsi$.  The result of
this straightforward exercise is,
\re{
\vB =  ((\h c)/2)^\hfs \pmdsq^{-\qrs}(\del\x(\vPsi + \vPsist)),
\cr\hbox{\ }\cr
\vE = i((\h c)/2)^\hfs \pmdsq^\qrs(\vPsi - \vPsist),
}{8e}
which can be checked by insertion of this result into Eq.~(8b), bear%
ing in mind the Eq.~(8c) reqirement of transverse free-photon spin.

We see that the source-free Maxwell equation $\del\dt\vB = 0$ holds
identically for the $\vB$ of Eq.~(8e), while the source-free Maxwell
equation $\del\dt\vE = 0$ follows from the $\vE$ of Eq.~(8e) in con%
junction with the Eq.~(8c) requirement of transverse free-photon spin.
  
The source-free Maxwell equation $\dot\vE = c(\del\x\vB)$ follows from
application of the Eq.~(8d) effective \Schr\ equation for the free
photon to the $\vE$ of Eq.~(8e) plus from taking the curl of the $\vB$
of Eq.~(8e) while bearing in mind the Eq.~(8c) requirement of trans%
verse free-photon spin.

The source-free Maxwell equation $\dot\vB =-c(\del\x\vE)$ follows from
application of the Eq.~(8d) effective \Schr\ equation for the free
photon to the $\vB$ of Eq.~(8e) plus from taking the curl of the $\vE$
of Eq.~(8e).

We thus see that the \e{quantum} free-photon effective \Schr\ equation
and transverse spin requirement both follow from the four \e{classi%
cal} source-free Maxwell equations, while those four \e{classical}
source-free Maxwell equations conversely follow from the \e{quantum}
free-photon effective \Schr\ equation and transverse spin requirement.
Therefore we here also have an archetypal quantum-classical complemen%
tary isomorphism.

It is certainly fascinating that an \e{isomorphic mapping} of the
mathematics which describes a \e{quantum \Schr-equation physical sys%
tem} (which here is the complex-valued transverse-vector free-photon
wave function) \e{can be so perfectly suited to the description of a
related classical complementary physical system} (which here is the
source-free classical electromagnetic field).
\vfil
\break
\noindent{\frtbf References}

\vskip 0.25\baselineskip

{\parindent = 15pt
\sk\item{[1]}
S. S. Schweber,
\e{An Introduction to Relativistic Quantum Field Theory}
(Harper \& Row, New York, 1961).
\sk\item{[2]}
S. K. Kauffmann,
arXiv:1210.7552 [physics.gen-ph] (2012).
\sk\item{[3]}
J. D. Bjorken and S. D. Drell,
\e{Relativistic Quantum Mechanics}
(McGraw-Hill, New York, 1964).
\sk\item{[4]}
Wikipedia,
``Complementarity (physics)'',
http://en.wikipedia.org/wiki/Complementarity\_(physics).
\sk\item{[5]}
S. K. Kauffmann,
arXiv:1101.0168 [physics.gen-ph] (2011).
\sk\item{[6]}
S. K. Kauffmann,
arXiv:1011.6578 [physics.gen-ph] (2010).
\sk\item{[7]}
J. D. Bjorken and S. D. Drell,
\e{Relativistic Quantum Fields}
(McGraw-Hill, New York, 1965).
}
\bye